\documentclass[a4paper,11pt]{article}

\pdfoutput=1 
  
\usepackage{jcappub} 
\usepackage[T1]{fontenc} 
\usepackage{amsthm}
\theoremstyle{definition} 
\newtheorem{exmp}{Example}[section]
\title{ \boldmath Chaotic inflation limits for non-minimal models with a Starobinsky attractor}
  
\author[a]{B. Mosk} 
\author[a]{J. P. van der Schaar}

\affiliation[a]{Delta Institute for Theoretical Physics \\ Institute of Physics, University of Amsterdam\\Science Park 904, Amsterdam, the Netherlands}

\emailAdd{b.mosk@uva.nl}
\emailAdd{J.P.vanderSchaar@uva.nl}

\abstract{We investigate inflationary attractor points by analysing non-minimally coupled single field inflation models in two opposite limits; the `flat' limit in which the first derivative of the conformal factor is small and the `steep' limit, in which the first derivative of the conformal factor is large. We consider a subset of models that yield Starobinsky inflation in the steep conformal factor, strong coupling, limit and demonstrate that they result in $\phi^{2n}$-chaotic inflation in the opposite flat, weak coupling, limit. The suppression of higher order powers of the inflaton field in the potential is shown to be related to the flatness condition on the conformal factor. We stress that the chaotic attractor behaviour in the weak coupling limit is of a different, less universal, character than the Starobinsky attractor.  Agreement with the COBE normalisation cannot be obtained in both attractor limits at the same time and in the chaotic attractor limit the scale of inflation depends on the details of the conformal factor, contrary to the strong coupling Starobinsky attractor.}
 
\begin{document}
\maketitle
\flushbottom

\section{Introduction}
\label{sec:intro}

To organise, and possibly increase, our understanding of inflationary models, as a starting point it would seem beneficial to classify the range of possibilities. It has been realised for quite some time that with the introduction of non-minimal couplings the space of phenomenologically acceptable theories of inflation can be expanded considerably \cite{Accetta:1985, Salopek:1988, Bezrukov:2007}, typically in fact ameliorating the required degree of fine-tuning \cite{Fakir:1990, Kaiser:1993}. Indeed,  the addition of a non-minimal coupling in general has the effect of stretching certain regions of  the potential, resulting in flat plateaus allowing for an extended inflationary slow-roll phase. The appearance of a non-minimal coupling is expected from the point of view of string theory or supergravity, due to the natural presence of dilaton and/or Kaluza-Klein scalars that mix with the gravitational sector. Moreover, a particular class of these non-minimal models that can (partially) be motivated from conformal supergravity theory, always lead to an inflationary phase that is Starobinsky like \cite{Starobinsky}. In other words, all these models reduce to the Starobinsky attractor in the strong non-minimal coupling limit \cite{Kallosh:2013lkr,Linde:2014nna}, see also \cite{Kallosh-mf, Kaiser-mf} for the multifield generalisation. This presents an interesting, economic and unified way of understanding (the predictions of) these models that in addition seems to be right in the sweetspot of the Planck data that seems to suggest a low tensor-to-scalar ratio \cite{Planck}\footnote{Another approach of interest leading to a (reduced) tensor-to-scalar ratio in the Planck sweetspot is warm inflation \cite{warminflation}}. Moreover, the so-called `induced inflation' subset of this class was shown to preserve perturbative unitarity up to the Planck scale \cite{Giudice:2014toa}, a basic first requirement for the internal consistency of these models. These models were recently also shown to have a natural embedding in no-scale supergravity models \cite{Pallis}.
 
Similarly, in a priori unrelated work it was pointed out that another class of non-minimal models reduce to chaotic models of inflation in a particular limit \cite{Lee:2014spa, Kallosh:2013maa, Kallosh:2013hoa, Kallosh:2014rga, Kallosh:2013yoa}. In view of the currently available cosmological data these attractor points (Starobinsky and chaotic \cite{Starobinsky, Linde:1983gd}) are of specific interest since they lie at opposite ends of the observational sweet spot in the spectral index $n_s$ versus tensor-to-scalar ratio $r$ plane \cite{Hinshaw:2012aka, Planck, BICEP2}. The appearance of a chaotic inflation limit might also provide some insights on the fine-tuning issues that plague effective field theories of chaotic inflation. Here we point out that a particular subset of models that feature a Starobinsky attractor in a strong coupling limit also appear to reduce to a chaotic attractor in the opposite weak coupling limit. We determine generic (sufficient) conditions for a chaotic model of inflation to appear in the weak coupling limit and parametrise the leading higher-order corrections. If the flat and steep limits can be controlled by a single flow parameter $\alpha$, like in examples (\ref{exmp1},\ref{exmp4}), then $\alpha$ parametrises a trajectory in the $(n_s,r)$-plane, connecting the Starobinsky attractor point with the chaotic attractor point. Notably, as a consequence the flow of these models seems to lie within the currently phenomenologically allowed range of predictions for $n_s$ and $r$. As we will point out however, the chaotic attractor is less universal in the sense that the predicted scale of inflation depends on the details of the conformal factor, contrary to the strong coupling Starobinsky attractor. 
 
The organisation of this paper is as follows. In section (\ref{sec:inm}) we define the `flat' and `steep' conformal limits in the context of single scalar field models that are non-minimally coupled to Einstein gravity.  In section (\ref{sec:si}), we identify under which conditions the flat conformal factor limit leads to slow-roll inflation and we explicitly show the appearance of chaotic models of inflation in this limit. In section (\ref{sec:fl}) we show that the weak coupling limit of a subset of induced inflation models effectively yields $\phi^2$-chaotic inflation. In section (\ref{sec:nm}) we further generalise and identify another set attractor points that correspond to $\phi^{2n}$-chaotic inflation. 

We note that similar results are reported in recent work by Roest, Kallosh and Linde \cite{roest:toappear}.

\section{Non-minimally coupled models and the Starobinsky attractor}
\label{sec:inm}
In this section we describe the framework of single scalar field models with non-minimal coupling to Einstein gravity, for which the Lagrangian is given by 
\begin{equation}\label{general}
\begin{aligned}
\mathcal{L} &= \sqrt{-g}\left(\frac{1}{2}\Omega(\phi)R-\frac{K(\phi)}{2}(\partial\phi)^2 -U(\phi) \right),
\end{aligned}
\end{equation}
where we require that at the vacuum configuration $\Omega\rightarrow 1$ and $U\rightarrow 0$\footnote{We have set the reduced Planck mass to one. The vanishing of the potential $U$ corresponds to a small cosmological constant. }. 

One can write (\ref{general}) in the Einstein frame with a canonically normalised kinetic energy term, by performing a Weyl transformation followed by a field redefinition. In appendix (\ref{sec:jtoe}) we give expressions for the Einstein frame potential slow-roll parameters in terms of the Jordan frame quantities $\Omega$ and $U$. These expressions greatly simplify if $K=\Omega$, which can be accomplished by doing an initial field redefinition. However, we will choose $K=1$, just to be consistent with the formulation of induced inflation in \cite{Giudice:2014toa}. After performing a Weyl transformation $g_{ab}\rightarrow \Omega^{-1}g_{ab}$, the Lagrangian reads:
\begin{equation}\label{generalk1}
 \begin{aligned}
  \mathcal{L} &= \sqrt{-g}\left(\frac{R}{2} -\frac{1}{2}\left(\frac{1}{\Omega}+\frac{3}{2}\left( \frac{\Omega'}{\Omega}\right)^2 \right)(\partial \phi)^2-\frac{U}{\Omega^2}\right),
 \end{aligned}
\end{equation}
where the prime denotes differentiation with respect to the scalar $\phi$. 
The kinetic energy term can be canonically normalised by performing a field redefinition $\chi(\phi)$ such that
\begin{equation}\label{redefinitionk1}
\begin{aligned}
 \left(\frac{\delta \chi}{\delta \phi}\right)^2 &= \frac{1}{\Omega}+\frac{3}{2}\left(\frac{\Omega'}{\Omega}\right)^2.
\end{aligned} 
\end{equation}
We will analyse these non-minimally coupled models in two limits, in which one of the two contributions to the Einstein frame kinetic term dominates (\ref{generalk1}). We will suggestively call these the \emph{flat} and \emph{steep} conformal factor limits respectively: \\ \\
\makebox{
\begin{minipage}[T]{0,4\textwidth}
\textit{flat limit} 
\begin{equation}\label{weak}
 \begin{aligned}
 \frac{3}{2}\frac{\Omega'^2}{\Omega} \ll 1
 \end{aligned}
\end{equation}
\end{minipage}

\begin{minipage}[T]{0,4\textwidth}
\textit{steep limit}
 \begin{equation}\label{strong}
 \begin{aligned}
 \frac{3}{2}\frac{\Omega'^2}{\Omega} \gg 1.
 \end{aligned}
\end{equation}
\end{minipage}
} \\ \\
In these limits, starting from the usual Einstein frame definitions for the inflationary slow-roll parameters \cite{Liddle:1994dx}(see appendix \ref{slowroll}), the expressions for the inflationary slow-roll parameter $\epsilon$ and $\eta$, in terms of $U$ and $\Omega$ (\ref{jepsilon},\ref{jeta}), simplify considerably. For the first order slow-roll parameter $\epsilon$ one arrives at the following expressions \\ \\
\makebox{
\begin{minipage}[T]{0,4\textwidth}
\textit{flat limit}
\begin{equation}
 \begin{aligned}
  \epsilon &\approx \frac{\Omega}{2}\left(\frac{U'}{U}-2\frac{\Omega'}{\Omega} \right)^2
 \end{aligned}
\end{equation}
\end{minipage}
 
\begin{minipage}[T]{0,4\textwidth}
\textit{steep limit}
 \begin{equation}
 \begin{aligned}
  \epsilon &\approx \frac{\Omega^2}{3 \Omega'^2} \, \left(\frac{U'}{U}-2 \frac{\Omega'}{\Omega} \right)^2.
 \end{aligned}
\end{equation}
\end{minipage}
} \\ \\
The slow-roll conditions can be naturally satisfied by Jordan frame potentials $U$ that are proportional to $\Omega^2$ up to terms higher order in the slow-roll approximation, for both the flat and steep limits. Requiring the inflationary model to be Starobinsky-like in the steep limit in fact determines the Jordan frame potential \cite{Giudice:2014toa} 
\begin{equation}\label{as}
U =\lambda  \,(\Omega-1)^2 \, .
\end{equation}
We will refer to these models as \emph{asymptotic Starobinsky models}. This particular relation between the conformal factor and the Jordan frame potential allows a further simplification of the slow-roll parameters in the flat and steep conformal factor limits  \\ \\ 
\makebox{
\begin{minipage}[T]{0,4\textwidth}
\textit{flat limit}
\begin{equation}
 \begin{aligned}
  \epsilon &\approx 2 \left( \frac{\Omega'^2}{\Omega} \right) \, \frac{1}{\left( \Omega - 1 \right)^2}   
  \end{aligned}
\end{equation}
\end{minipage}

\begin{minipage}[T]{0,4\textwidth}
\textit{steep limit}
 \begin{equation}
 \begin{aligned}
 \epsilon &\approx \frac{4}{3} \, \frac{1}{\left( \Omega - 1 \right)^2} \, .
  \end{aligned}
\end{equation}
\end{minipage}
} \\ \\

It is important to note, as should also be clear from the above expressions,  that the flat or steep conformal factor limits do not necessarily imply the slow-roll conditions. They do seem to be sufficient to allow for regions in field space where the slow-roll conditions are met. From now on we will mostly be interested in considering the (opposite) flat conformal factor limit of the asymptotic Starobinsky models. First we identify sufficient conditions for the conformal factor such that a slow-roll inflationary regime exists in the flat conformal factor limit and, if the flat limit exists, to what type of inflationary model this leads. 
   
\section{The flat conformal factor limit}
\label{sec:si}
In this section we will investigate the flat conformal factor limit of asymptotic Starobinsky models. We will show that a generic power law implementation of the flat conformal factor limit is sufficient to make sure that a slow-roll limit can be satisfied in some region of field space. If we are interested in considering a flat conformal factor limit (\ref{weak}) that can be satisfied over a large enough field range, a natural procedure would be to constrain all derivatives of $\Omega$ around the vacuum field value $\phi_{\text{vac}}$ to be sufficiently small. Note that for the asymptotic Starobinsky models that we are considering the vacuum field value is defined by $U(\phi_{\text{vac}})=0$. It is straightforward to check that this imposes the following condition on the derivatives of $\Omega$ at $\phi_{\text{vac}}$, denoted by $\Omega^{(n)}_{\text{vac}}$
\begin{equation}\label{finetuning}
\Omega^{(n)}_{\text{vac}} = O\left((\alpha)^n\right) \, ,
\end{equation}
for some small parameter $\alpha$, such that $\alpha (\phi -\phi_{\text{vac}}) \ll 1$, with $\phi_{\text{vac}}<\phi<\phi_N$ where $\phi_N$ denotes the field value to allow for $N$ e-folds of slow-roll inflation. In general it is useful to perform a shift to the field variable $\tilde{\phi}$ defined as $\tilde{\phi}=\phi-\phi_{\text{vac}}$. After the shift the same conditions (\ref{finetuning}) apply, with $\phi$ replaced by $\tilde{\phi}$ and $\tilde{\phi}_{\text{vac}}=0$. These conditions on the derivatives of $\Omega$ at $\phi_{\text{vac}}$ resemble the slow-roll conditions and as such could be considered as fine-tuning. When we discuss specific examples of the flat conformal factor limit we will come back to this point. 

An expansion of the Einstein frame potential in terms of the canonically normalised field $\chi$ (\ref{redefinitionk1}) around $\chi_{\text{vac}}=0$ explicitly shows the relation between the flat conformal factor condition (\ref{finetuning}) and the suppression of higher order powers of $\chi$, for the asymptotic Starobinsky model $U= \lambda(\Omega-1)^2$
\begin{equation}\label{vexpansion}
 \begin{aligned}
   \left. \frac{U}{\Omega^2}\right|_{\phi = \phi(\chi)} &=  \lambda \Omega'^2_{\text{vac}}\chi^2+  \lambda\left(\Omega'_{\text{vac}}\Omega''_{\text{vac}}-\frac{3}{2}\Omega'^3_{\text{vac}}\right)\chi^3+\\
  &+ \frac{\lambda}{24} \left(-50\Omega'^2_{\text{vac}}\Omega''_{\text{vac}}+\frac{75}{2}\Omega'^4+6\Omega''^2_{\text{vac}}+8\Omega'_{\text{vac}}\Omega'''_{\text{vac}}\right)\chi^4+\dots \, , 
 \end{aligned}
\end{equation} 
where it should be understood that the dots not only include higher powers of $\chi$ but also corrections to the coefficients higher order in the flat conformal factor limit (\ref{vfullexpansion}). So we conclude that the leading term in the power law expansion is the $\chi^2$-term, as long as $\Omega'_{\text{vac}} \neq 0$. This is of course recognised as the potential for (quadratic) chaotic inflation. Note that the slow-roll conditions are violated for small $\chi$, so the field range where the slow-roll conditions apply is smaller than the field range where the flat conformal factor conditions apply. Higher order terms are polynomially suppressed by virtue of the flat conformal factor condition (\ref{finetuning}) on the higher order derivatives of $\Omega$. Smallness of the higher order terms in the Einstein frame potential of a $\chi^2$-chaotic inflation model can be interpreted as the smallness of the variation of the conformal factor $\Omega$ in the Jordan frame of a Lagrangian (\ref{general}).
 
From the above expansion we also see that when $\Omega'_{\text{vac}} = 0$, but $\Omega''_{\text{vac}}\neq 0$, then the first nonzero term in expansion (\ref{vexpansion}) is the $\chi^4$-term. The higher order terms are again suppressed by virtue of the flat conformal factor condition (\ref{finetuning}). The first non-zero derivative of $\Omega$ therefore determines the (higher-order) model of chaotic inflation. Note that although the coefficients are different for different $\Omega$ (and $\lambda$), the slow-roll parameters for chaotic models do not depend on the coefficients and as such the predictions in the $n_s$ versus $r$ plane will be the same, as we will soon show explicitly. Of course, the scale of inflation is related to the specific value of the coefficient. To agree with observational constraints, for quadratic chaotic inflation the COBE normalisation implies that the mass parameter should roughly equal $10^{-5}$ (in natural units). For a given $\lambda$ this further constrains the first derivative of the conformal factor. So although these models all give the same predictions for $n_s$ and $r$, they are observationally distinguished in their prediction for the magnitude of the density perturbations. This is different from the steep conformal factor limit, where the scale of inflation is uniquely determined by the parameter $\lambda$. This perhaps favours fixing $\lambda$ to a value that agrees with the COBE normalisation in the Starobinsky limit, but as should be clear from the above discussion this will then not reproduce the COBE normalisation in the weak non-minimal coupling limit.  

After this general discussion, let us now move on to the general expressions for the slow-roll parameters in this limit and provide some specific examples in which a non-minimal coupling parameter governs a flow between the flat and steep conformal factor limits.

\section{Chaotic fixed points for asymptotic Starobinsky models}
\label{sec:fl}
In the previous section we saw that in general, when the flat conformal factor condition is satisfied, the Einstein frame potential will be that of chaotic inflation. Here we will first determine the consequences of this general result for the first and second slow-roll parameters, explicitly using the flat conformal factor condition (\ref{finetuning}). Subsequently we give some specific examples in which this behaviour is realised. 

If the conformal factor $\Omega$ satisfies the flat conformal factor limit (\ref{finetuning}), we can expand $\Omega$ as
\begin{equation}\label{alphaexpansion}
\begin{aligned}
 \Omega &= 1+\sum_{m=1} \Omega_m \alpha^m \tilde{\phi}^m,
\end{aligned}
\end{equation}
where we extracted the coefficients $\Omega_m$ that are of order $O(1)$ and again introduced the small parameter $\alpha$ that should satisfy $\alpha \tilde{\phi} \ll 1$ for $0 < \tilde{\phi} < \tilde{\phi}_N$.

In appendix (\ref{slowroll}) we compute the slow-roll parameters for several different cases. If $\Omega_1 \neq 0$, we find (\ref{gso1o2} with $n=1$):
\begin{equation}\label{o1o2}
\begin{aligned}
  \epsilon &= \frac{2}{\tilde{\phi}^2}+O\left(\alpha\tilde{\phi}\right), \ \ \
  \eta = \frac{2}{\tilde{\phi}^2} +O\left(\alpha\tilde{\phi}\right),
\end{aligned}
\end{equation}
which indeed corresponds to leading order to the results of $\phi^2$-chaotic inflation. One could imagine imposing the condition that $\Omega$ is an even function. In that case, for $\Omega_2 \neq 0$, we find (\ref{symmetric} with $n=1$)
\begin{equation}
\begin{aligned}
  \epsilon &= \frac{8}{\tilde{\phi}^2}+O\left((\alpha\tilde{\phi})^2\right)  , \ \ \ \ \
  \eta = \frac{12}{\tilde{\phi}^2} +O\left((\alpha\tilde{\phi})^2\right),
\end{aligned}
\end{equation} 
which corresponds to leading order to the results of $\phi^4$-chaotic inflation. The explicit expressions for the subleading parts can be found in appendix (\ref{slowroll}). \\

To illustrate this further and relate the flat and steep conformal limits to a continuous non-minimal coupling parameter to be able to consider the flow behaviour as a function of this non-minimal coupling, let us give two examples where the conformal factor naturally satisfies the flatness limit (\ref{finetuning}). In the first example we analyse induced inflation models. This agrees with the analysis done in \cite{roest:toappear}. 

\begin{exmp}\label{exmp1}\textbf{Induced inflation} \\
\emph{Induced inflation}\footnote{Originally these models were studied as examples where the spontaneous symmetry breaking in (non-minimal) induced gravity models would allow for slow-roll inflation \cite{Accetta:1985}. More recently it was pointed out that the Einstein frame potential does not include power series in terms of $\xi$ in the large $\xi$ limit. Hence perturbative unitarity is not violated before reaching the Planck scale in models of induced inflation \cite{Giudice:2014toa}.} is a particular subset of asymptotic Starobinsky models, with $\Omega(\phi) = \xi f(\phi)$ and $f(0)=0$, where $\xi$ is a coupling parameter. \\

\noindent \textbf{Monomial: $\Omega = \xi \phi^m$} \\ 
Since we demand that $\Omega(\phi_{\text{vac}}) =1$, we find $\phi_{\text{vac}} = \xi^{-\frac{1}{m}}$. Defining $\tilde{\phi} = \phi-\xi^{-\frac{1}{m}}$ one can write the following expansion of $\Omega$ in terms of $\tilde{\phi}$:
\begin{equation} \label{iiexpo}
 \begin{aligned}
  \Omega &= \xi \phi^m \\
  &= \xi \left(\tilde{\phi}+\xi^{-\frac{1}{m}} \right)^m \\
  &= 1 +  \sum_{i=1}^m \tilde{\phi}^i\xi^{\frac{i}{m}}\binom{m}{i}.
 \end{aligned}
\end{equation}
Expansion (\ref{iiexpo}) explicitly shows that induced inflation with $f=\phi^m$ provides a realisation of the flat conformal factor limit (\ref{finetuning}) with the identification $\alpha \sim \xi^{\frac{1}{m}}$ for small $\xi$. Induced inflation with a monomial conformal factor \emph{also} guarantees that $\Omega'_{\text{vac}}\neq 0$. This means that (\ref{iiexpo}) is of the form (\ref{alphaexpansion}) with $\Omega_1 = m$ and $\Omega_2 = \binom{m}{2}$ and identifying $\xi^{\frac{1}{m}}\sim \alpha$ (see also example \ref{o1o2exp}). Hence for small $\xi$ the result (\ref{o1o2}) applies; to leading order this model corresponds to quadratic chaotic inflation. 
\begin{figure}[tbp]
\centering 
\includegraphics[width=.4\textwidth]{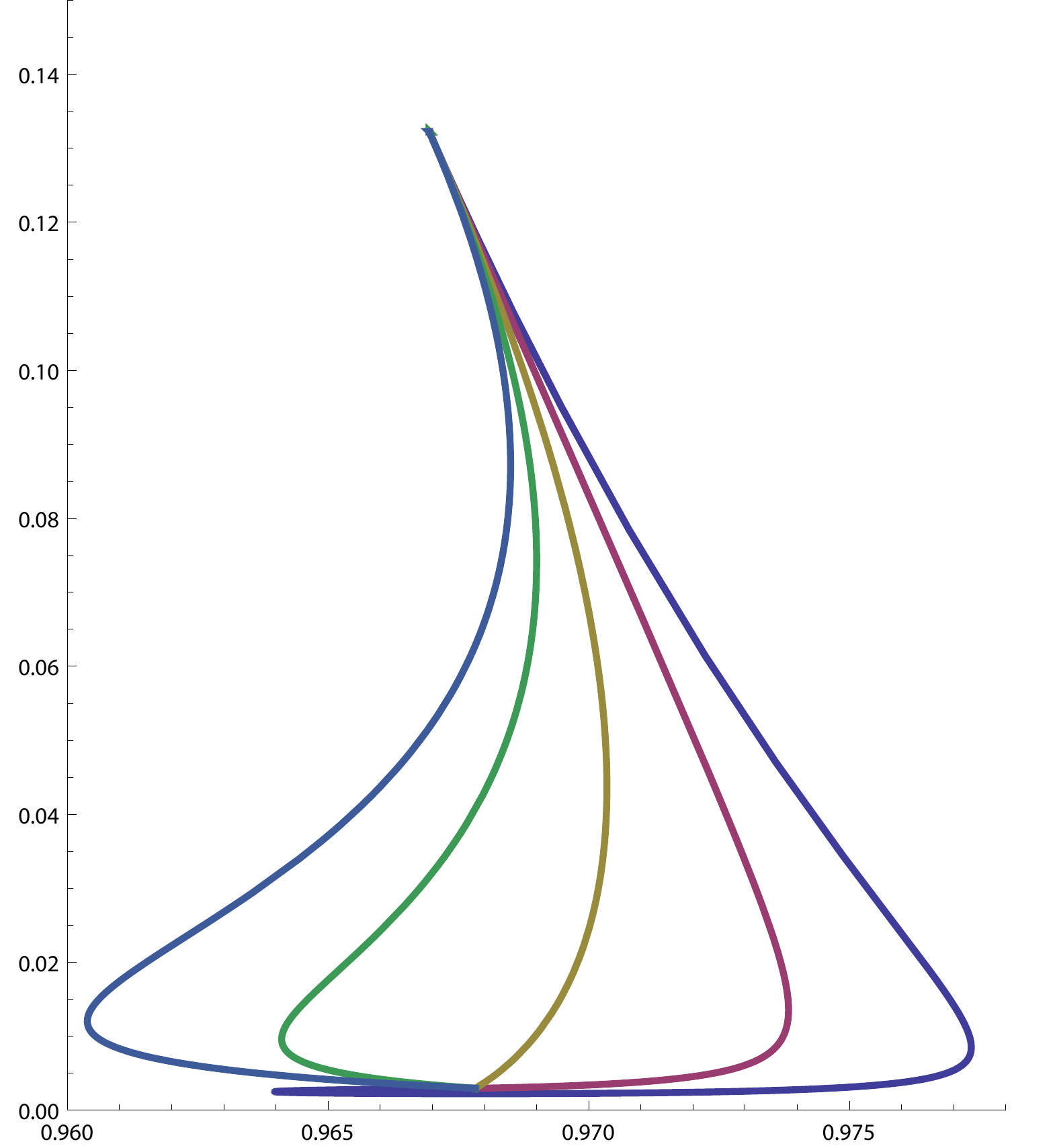}
\hfill
\includegraphics[width=.4\textwidth]{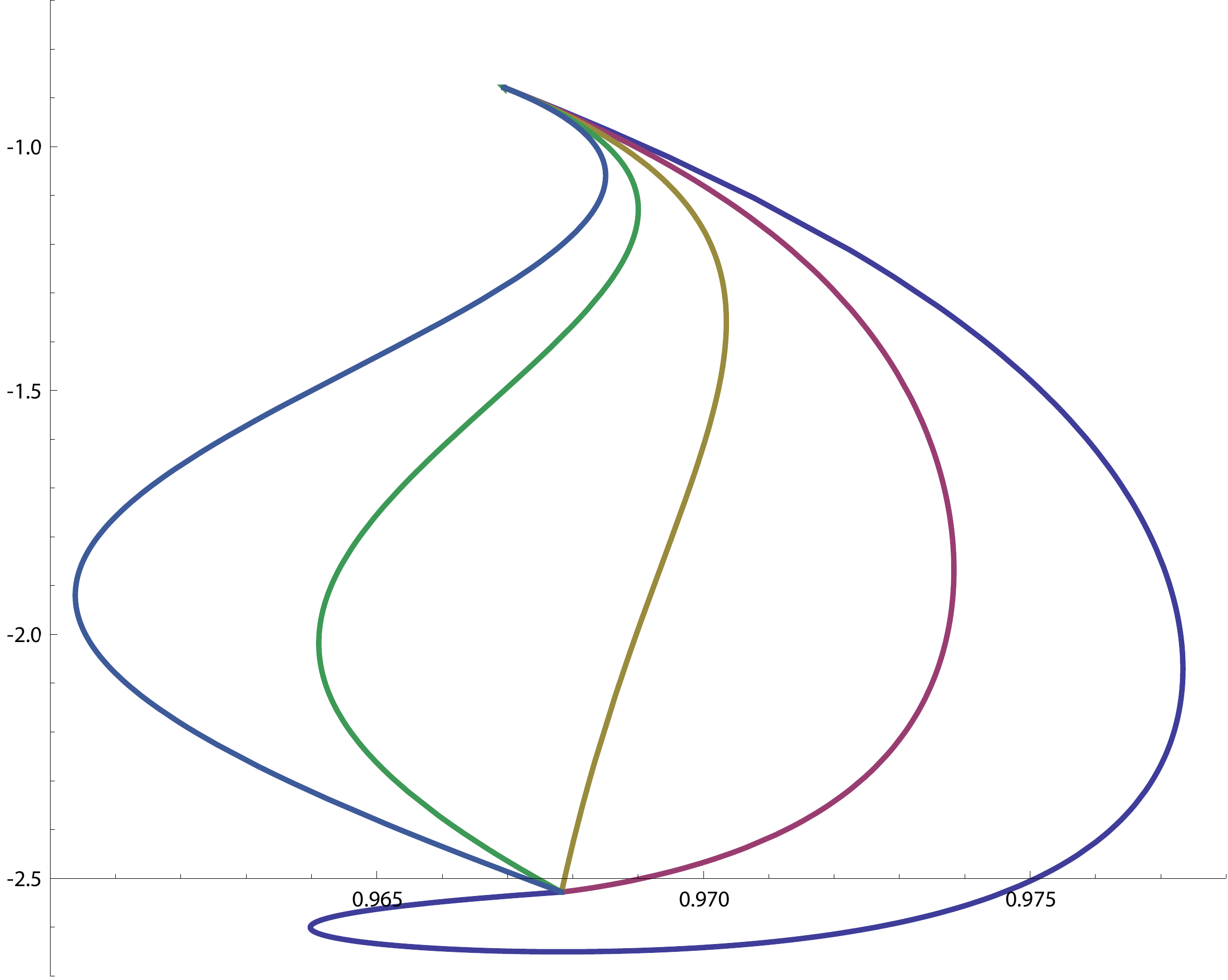}
\caption{\label{fig:IIphi2} $(n_s,r)$-plane with linear and $\log_{10}$-scale on the vertical $r$-axis, for monomial induced inflation (\ref{iiexpo}) with $m=0.5,  1, 2, \ 4, \ 8$, from right to left.}
\end{figure}
We can verify this directly by computing the Einstein frame potential $V(\chi)$
  \begin{equation}\label{m23}
  \begin{aligned}
V &= \lambda m^2\xi^{\frac{2}{m}}\chi^2+O(\xi^{\frac{3}{m}}) .
  \end{aligned}
 \end{equation}
The slow-roll parameters are independent of the mass parameter $M^2= 2 \lambda \, m^2 \, \xi^{\frac{2}{m}}$ of this chaotic inflationary potential, but the mass does determine the magnitude of the density perturbations, which in natural units should roughly equal $10^{-5}$ to be in agreement with the COBE normalisation. The coupling $\xi$ parametrises a trajectory in the $(n_s,r)$-plane, connecting the Starobinsky attractor point with the chaotic attractor point (see figure \ref{fig:IIphi2}). If one fixes the parameter $\lambda$ along the flow, then the prediction for the magnitude of the (scalar) density perturbations will not be in agreement with observation in the strict weak (non-minimal) coupling limit $\xi \rightarrow 0$. Alternatively, one could introduce a rescaled coupling $\tilde{\lambda}$ killing off the $\xi$ dependence. However, one would then have a similar problem in the strong coupling limit and moreover, the required rescaling would be different for different values of $m$. In contrast, the normalisation in the opposite strong coupling limit is independent of the power $m$ of the monomial.  

In addition to the properties of the fixed points, in figure \ref{fig:IIphi2} the flow as a function of the non-minimal coupling $\xi$ is plotted. A notable feature, in contrast to the strong coupling Starobinsky attractor, is that the approach to the weak coupling chaotic fixed point is clearly seen to be universal (independent of $m$), as can be analytically confirmed by determining the first order corrections in $\xi$ around the chaotic fixed point. \\

\noindent \textbf{Finite polynomial: $\Omega = \xi \left( \dots +\phi^m\right)$} \\
A minimal extension of the monomial conformal factor is to consider the case where $f$ is a polynomial in $\phi$ with a finite number of terms. In this case, the highest power of $\phi$ determines the vacuum value of the field $\phi_{\text{vac}}$; if $f(\phi) = \dots+\phi^m$, where the dots indicate lower powers of $\phi$, then the previous analysis goes through, with 
\begin{equation}\label{polexp}
\begin{aligned}
 \phi_{\text{vac}} &= \xi^{-\frac{1}{m}}+O(\xi^0) \\
 \Omega &= 1 + m \xi^{\frac{1}{m}}\tilde{\phi}+O(\xi^{\frac{2}{m}})
 \end{aligned}
\end{equation}
and we arrive at the same conclusion: the weak coupling limit is described by quadratic chaotic inflation. Lower powers of $\phi$ in the polynomial do affect the subleading terms in expansion (\ref{polexp}). The coupling parameter $\xi$ parametrises a curve in the $(n_s,r)$-plane, connecting the strong coupling Starobinsky attractor point with the weak coupling chaotic attractor point. 

In an infinite series expansion of the conformal factor (as could be generated by quantum corrections) this sensitivity to the highest order term might be considered problematic. Neglecting that for now, it does imply that \emph{all induced inflation models with a finite number of powers of the field $\phi$} reduce to chaotic inflation in the weak coupling limit. \\ 

\noindent \textbf{Exponential: $\Omega = \xi f = \xi \left(e^{\beta \phi}-1\right)$ } \\
In the previous example we observed that $\Omega$ is sensitive to the highest power in the expansion of $f$, in the weak coupling limit. In this example we investigate whether the weak coupling limit of induced inflation still leads to $\phi^2$-chaotic inflation if $f$ is an infinite series in $\phi$. If $f =  \left(e^{\beta \phi}-1\right)$, then $\phi_{\text{vac}} = \frac{1}{\beta}\ln\left(1+\frac{1}{\xi}\right)$ and $\Omega = e^{\beta \tilde{\phi}}(\xi+1)-\xi$. For large $\xi$ this model \emph{will} satisfy the steep limit (\ref{strong}) and hence have the Starobinsky model as a strong coupling attractor point. For very small $\xi$ this model does generally not satisfy the flat limit (\ref{weak}), unless $\beta$ is tuned for this purpose. So in general the $\xi\rightarrow 0$ limit does \emph{not} correspond to quadratic chaotic inflation. \\

In fact the above conclusion that exponential functions generically do not feature chaotic inflation attractors can be changed by making a different identification of the coupling parameter. Below we briefly present this case as another example.\\ 

\begin{exmp}\label{exmp4}
\textbf{$\Omega = e^{\xi \phi^m}$ in weak coupling limit} \\ \\
The conformal factor $\Omega = e^{\xi \phi^m}$ naturally satisfies the flat conformal factor condition (\ref{finetuning}) for small $\xi$\footnote{An expansion of the kinetic term in the Einstein frame, at strong coupling, suggests that perturbative unitarity is violated before reaching the Planck scale for $m >1$, unlike induced inflation. }. We require that in the vacuum configuration, $\Omega =1$; this means that $\phi_{\text{vac}}=0$. In the $\xi \rightarrow 0$ limit we expand $\Omega$ in orders of $\xi$:
\begin{equation}\label{eeexpans}
 \begin{aligned}
  \Omega &= e^{\xi \phi^m} \\
  &= 1+\xi \phi^m+O(\xi^2)
 \end{aligned}
\end{equation} 
If $m=1$, then $\Omega_1 \neq 0$ and we directly recover the result (\ref{o1o2}), corresponding to $\phi^2$-chaotic inflation in the $\xi\rightarrow 0$ limit. For general $m$, we find (see example \ref{uaplus}):
 \begin{equation}
 \begin{aligned}
N &= \frac{\phi_N^2}{4m}+O(\xi) && \\
  \epsilon &=  \frac{2m^2}{\phi_N^2}+O(\xi) &&\approx \frac{m}{2N} \\
\eta &= \frac{2m(2m-1)}{\phi_N^2}+O(\xi) &&\approx \frac{2m-1}{2N}
\end{aligned}
\end{equation}
These results correspond to those of $\phi^{2m}$-chaotic inflation.\footnote{Adding extra terms in the exponent, e.g. $e^{\xi(\phi^n+\beta \phi^{n+m})}$ would explicitly violate the flat conformal factor condition (\ref{finetuning}).} \\
\end{exmp}
\end{exmp}

To conclude, we have confirmed that the weak coupling limit of some, but not all, models of induced inflation imply the flat conformal factor limit (\ref{finetuning}), yielding chaotic inflation attractor points. We also explicitly confirmed that the Einstein frame potential in this limit depends on the details of the conformal factor, implying that although the predictions for the spectral index and tensor-to-scalar ratio are the same for all these models and as such denote a fixed point, the magnitude of the density perturbations will depend on the details of the function $\Omega$ under consideration. In the opposite strong coupling limit, for the Starobinsky attractor, the magnitude of the density perturbations is instead independent of the details of the function $\Omega$, which can be used to uniquely fix $\lambda$ to agree with the COBE normalisation in the Starobinsky fixed point. In that sense the strong coupling Starobinsky fixed point can be considered more universal. 

\section{Generalised asymptotic Starobinsky models}
\label{sec:nm}
A straightforward extension of the asymptotic Starobinsky models (\ref{as}) with $U \propto \Omega^2(1-\Omega^{-1})^2$ is given by the set of Jordan frame potentials:
\begin{equation}\label{npotential}
 \begin{aligned} 
  U &= \lambda \Omega^2(1-\frac{1}{\Omega})^{2n}.
 \end{aligned}
\end{equation}
We will show that the steep conformal factor limit again corresponds to Starobinsky inflation, to leading order. We will also point out that the flat conformal factor limit corresponds to $\phi^{2n}$-chaotic inflation \footnote{In this subsection we will assume $\Omega'_{\text{vac}}\neq 0$ for simplicity}.

\paragraph{Steep conformal factor limit} 

In this limit (\ref{strong}) the field redefinition(\ref{redefinitionk1}) simplifies to:
\begin{equation}
\begin{aligned}
 \left(\frac{\delta \chi}{\delta \phi}\right)^2 &= \frac{3}{2}\left(\frac{\Omega'}{\Omega}\right)^2 &&\Rightarrow \Omega(\phi(\chi)) = e^{\pm\sqrt{\frac{2}{3}}\chi}.
 \end{aligned}
\end{equation}
The Einstein frame potential that corresponds to the potential (\ref{npotential}) in terms of $\chi$ now reads
\begin{equation}
 \begin{aligned}
  V(\chi) &= \lambda \left(1-e^{-\sqrt{\frac{2}{3}}\chi} \right)^{2n}.
 \end{aligned}
\end{equation}
For $n=1$, we recognise the Starobinsky potential. For general $n$ we have a generalised Starobinsky model which also has,
to leading order in $N$, for \emph{all} $n$\footnote{For large $n$ of order $N$ the slow-roll conditions can be violated.}:
\begin{equation}
 \epsilon \approx \frac{3}{4N^2}, \ \ \ \eta \approx -\frac{1}{N},
\end{equation}
similar to the standard Starobinsky model. In terms of the position in the $(n_s,r)$-plane of the strong coupling limit of these models, the attractor point coincides with that of Starobinsky inflation for all $n$, at least to leading order in $N$. The parameter $\lambda$ should be fixed to agree with the COBE normalisation. Requiring $60$ e-folds implies that $\epsilon \sim 10^{-3}$ and as a consequence $\lambda$ is fixed to roughly equal $10^{-10}$ (in natural units), i.e. a very small number. A value for $\lambda$ this small causes an obvious problem in the opposite weak coupling limit, where the potential depends on the non-minimal coupling $\xi$ and $\lambda$ and as a consequence, for small $\xi$, the predicted magnitude for scalar density perturbations will be too small.  

\paragraph{The flat conformal factor limit} 

In this limit (\ref{weak}), the first and second order slow-roll parameters are given by (see appendix \ref{slowroll}): 
\begin{equation}\label{epsilonweak}
  \begin{aligned}
   \epsilon &\approx  2n^2\frac{\Omega'^2}{\Omega}  \frac{1}{(\Omega-1)^2} \\
   \eta &\approx  \frac{\Omega'^2}{\Omega}\frac{2n(2n-1)}{(\Omega-1)^2}+\frac{2n}{\Omega-1}\left(\frac{\Omega''}{\Omega}-\frac{3}{2}\frac{\Omega'^2}{\Omega}\right).
  \end{aligned}
 \end{equation}
Given that $\Omega$ satisfies condition (\ref{finetuning}), we find to leading order in $N$\\ \\
\makebox{
\begin{minipage}[T]{0,4\textwidth}
\begin{equation}
 \begin{aligned}
  \epsilon &\approx \frac{2n^2}{\tilde{\phi^2}_N} \approx \frac{n}{2N}
   \end{aligned}
\end{equation}
\end{minipage} 
\begin{minipage}[T]{0,5\textwidth}
\begin{equation}
 \begin{aligned}
  \eta &\approx \frac{2n(2n-1)}{\tilde{\phi}_N^2} \approx \frac{2n-1}{2N}.
 \end{aligned}
\end{equation}
\end{minipage} 
}\\ \\

\begin{figure}[h]
\centering 
\includegraphics[width=.5\textwidth]{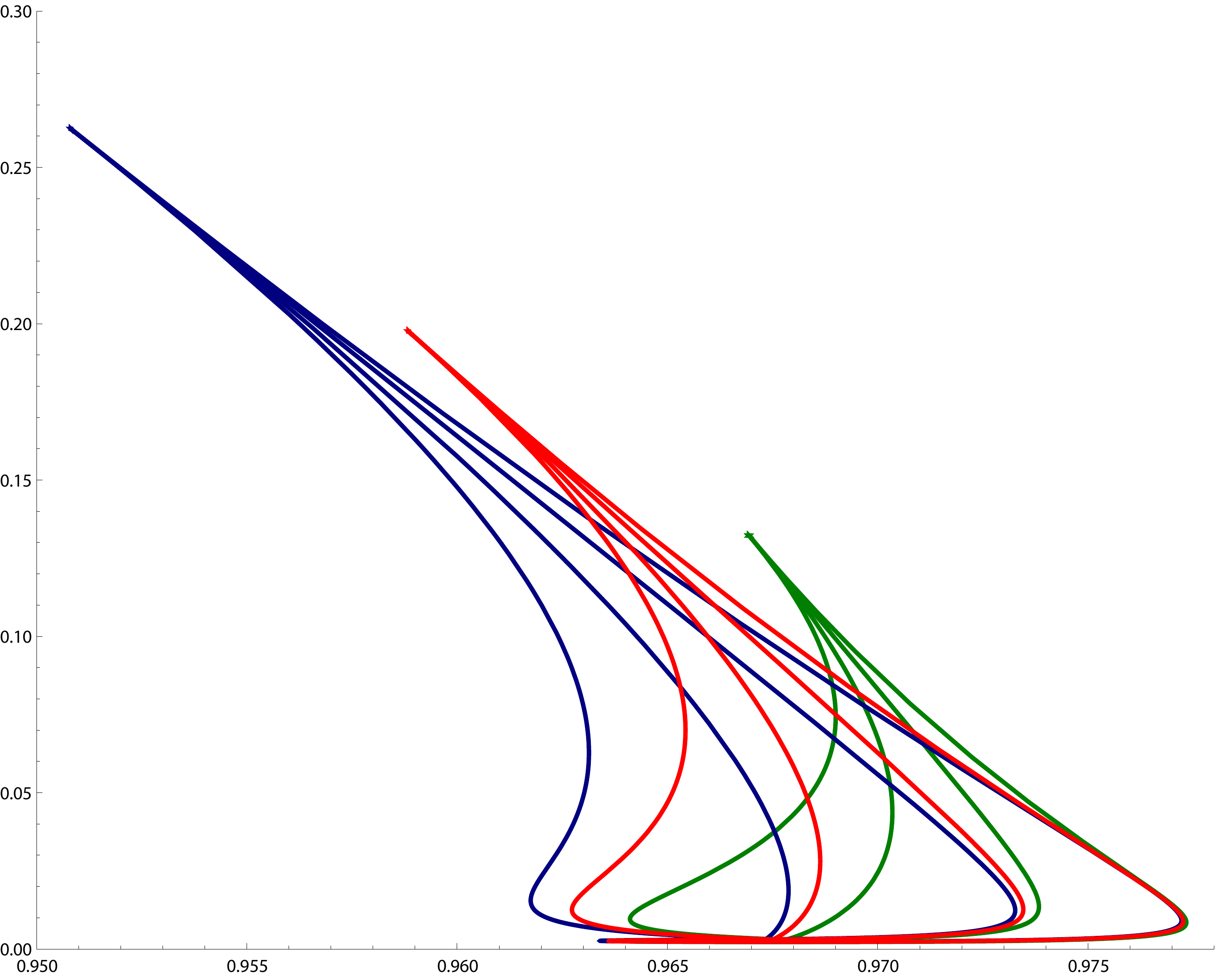}
\hfill
\includegraphics[width=.4\textwidth]{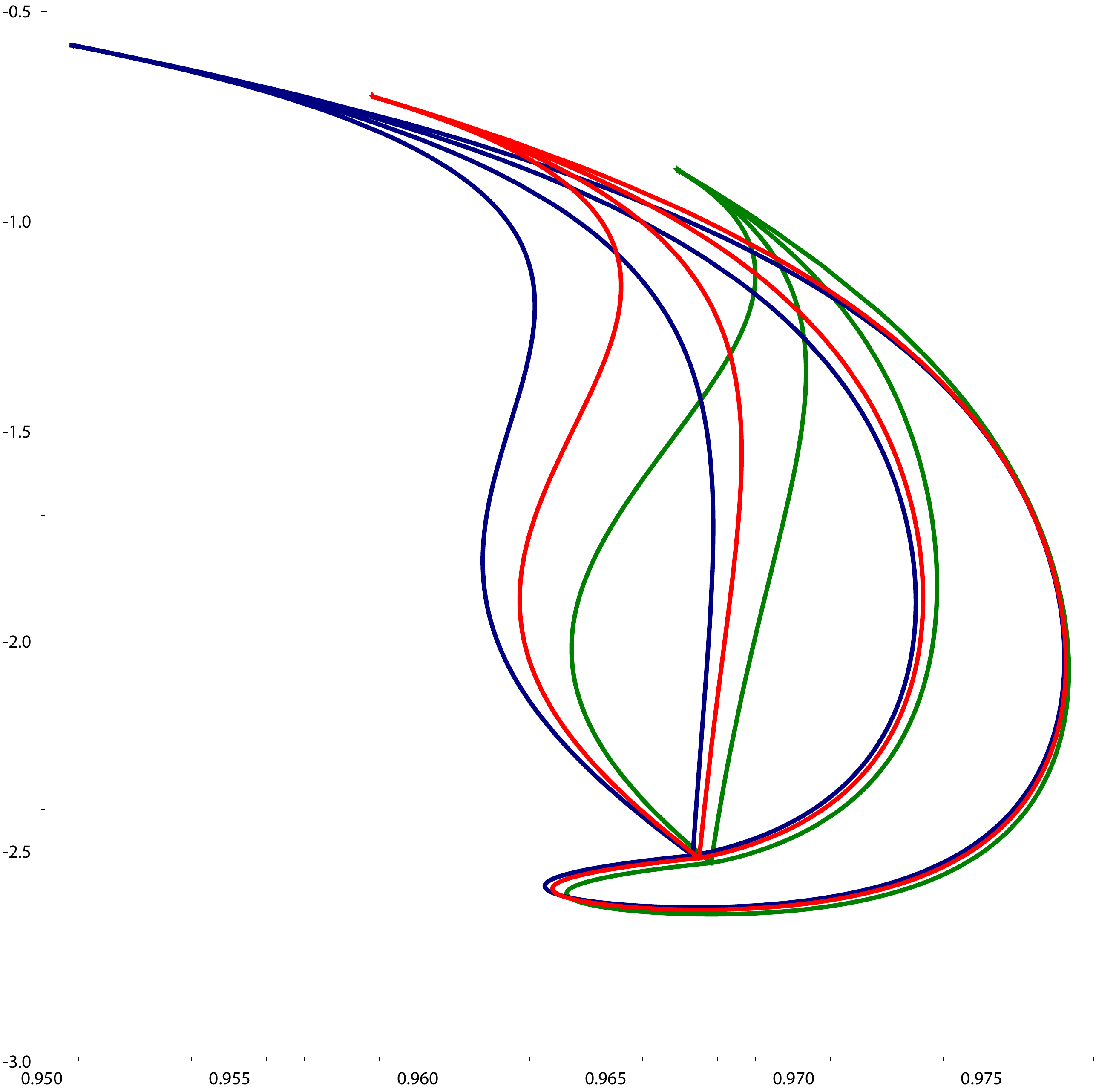}
\caption{\label{fig:IIphi4phi3phi2} $(n_s,r)$-plane with linear and $\log_{10}$-scale on the vertical $r$-axis, for generalised asymptotic Starobinsky models (\ref{npotential}) with $n = 1 ,\ 1.5 ,\ 2$ and $\Omega = \xi \phi^m$ (\ref{iiexpo}) with $m=0.5,  1, 2, \ 4$ for each $n$, from right to left. The large coupling limit corresponds to $\phi^2,\phi^3$ and $\phi^4$-chaotic inflation respectively.} 
\end{figure}

Indeed, for $n=1$ we recover $\phi^2$-chaotic inflation. In fact, these results correspond to $\phi^{2n}$-chaotic inflation for all $n$ (see figure \ref{fig:IIphi4phi3phi2}). Chaotic inflation models are fixed points for flat $\Omega$, whereas Starobinsky inflation is obtained in the opposite steep conformal factor limit. Looking at figure \ref{fig:IIphi4phi3phi2}, where the flow between fixed points in the $(n_s, r)$ plane is plotted for different values of the parameters $m$ and $n$, with $\Omega = \xi \phi^m$.

\paragraph{Approaching the chaotic attractor point}\label{angle} 
 
If the flat and steep limits can be controlled by a single flow parameter $\alpha$, like in examples (\ref{exmp1},\ref{exmp4}), then $\alpha$ parametrises a trajectory in the $(n_s,r)$-plane, connecting the Starobinsky attractor point with the chaotic attractor point. The analysis of induced inflation (see figures \ref{fig:IIphi2} and \ref{fig:IIphi4phi3phi2}) suggests that the approach of the chaotic attractor point is well behaved and along a certain universal angle, whereas the approach of the Starobinsky attractor point is more chaotic. The slope of the line in the $(n_s,r)$-plane close to the chaotic attractor point can be determined analytically by dividing $\frac{d r}{d \alpha}$ by $\frac{d n_s}{d \alpha}$. For polynomial induced inflation and polynomial universal attractor inflation, we find a slope of $\frac{-16}{2-\frac{1}{n}}$ for generalised asymptotically Starobinsky models (\ref{npotential})\footnote{This is in agreement with  \cite{Kallosh:2013tua}, where $(n=1)$}. This suggests that the slope does not depend on the details of $\Omega$, and only on the number $n$ in the potential (\ref{npotential}). However, there are explicit counterexamples\footnote{For example, consider $\Omega = 1 + \xi^2\phi^2+\xi^3\phi^3$}.
 
\section{Conclusions}
In this work we studied non-minimally coupled single scalar field inflation models with a Starobinsky attractor point in the strong coupling limit. We have identified the relevant  conditions on the conformal factor, corresponding to the flat and steep conformal factor limits, and shown how these can be obtained introducing a continuous coupling parameter, for instance in the context of induced inflation models. General (sufficient) conditions were determined that produce chaotic models of inflation in the flat conformal factor limit. Employing these general results we have confirmed the existence of chaotic fixed points for a subset of models that are asymptotically Starobinsky. As long as the first derivative of the conformal factor is non-zero the fixed point corresponds to the simplest quadratic model of chaotic inflation. The fine tuning of higher order powers in the potential of chaotic inflation was shown to be directly related to the flatness condition of the conformal factor. We also introduced and studied a straightforward generalisation of asymptotic Starobinsky models, parametrised by a power $n$, that reduces at weak coupling to a chaotic inflation fixed point of order $n$ (the leading power of the canonical Einstein scalar field $\chi$).  

One important observation is that this fixed point behaviour differs from the Starobinsky fixed point at strong coupling 
due to the explicit dependence of the mass (or the couplings in higher order chaotic models) on the details of the model under consideration. This means it should be considered less universal in the sense that different models in this class, although they all reduce to the same (chaotic) slow-roll parameters in the weak coupling limit, predict different scales of inflation. A related consequence is that the COBE normalisation cannot be matched in both fixed points at the same time. This observation is best illustrated in a three-dimensional flow plot that would include the predicted magnitude of density perturbations as a function of the non-minimal coupling $\xi$. By appropriately rescaling the coupling $\lambda$ to allow for a finite magnitude in the weak coupling limit, one would find that the weak coupling limit is not a fixed point in this $3$-dimensional space of inflationary parameters, whereas the strong coupling Starobinsky attractor remains a true fixed point. 

An important and interesting avenue for future work would be to better understand the UV embedding and (effective field theoretical) consistency of these classes of non-minimally coupled models, in particular from the point of view of string theory. We hope to come back to this issue in future work. 
  
 \acknowledgments
This work was initiated and influenced by a seminar given by D. Roest (Leuven, May 2014) who presented preliminary results obtained in a collaboration with R. Kallosh and A. Linde on the existence of a chaotic inflation attractor in the the weak coupling limit of induced inflation \cite{roest:toappear}. We thank D. Roest for very useful discussions and suggestions. This work is supported in part by the research program of the Foundation for Fundamental Research on Matter (FOM), which is part of the Netherlands Organisation for Scientific Research (NWO). This work is part of the D-ITP consortium, a program of the Netherlands Organisation for Scientific Research (NWO) that is funded by the Dutch Ministry of Education, Culture and Science (OCW).\\ \\
  
\appendix
\section{From Jordan frame to Einstein frame}
\label{sec:jtoe}
\subsection{Lagrangians}
A general Lagrangian of a theory with a non-minimally coupled scalar field can be written as:
  
\begin{equation}\label{jframe}
\begin{aligned}
\mathcal{L} &= \sqrt{-g}\left(\frac{1}{2}\Omega(\phi)R-\frac{K(\phi)}{2}(\partial\phi)^2 -U(\phi) \right).
\end{aligned}
\end{equation}
One could set the kinetic term $K(\phi)$ to a preferred functional $\tilde{K}(\chi(\phi))$ by doing a field redefinition $\phi \rightarrow \chi$ such that $\tilde{K}(\chi)\left(\frac{\delta \chi}{\delta \phi}\right)^2= K(\phi)$.

A Weyl transformation $g_{ab} \rightarrow \frac{1}{\Omega}g_{ab}$ can be used to bring the Lagrangian (\ref{jframe}) into the Einstein frame. The curvature scalar transforms in four spacetime dimensions in the following way under this transformation:
\begin{equation}\label{rtransform}
\begin{aligned}
R &\rightarrow \Omega \left(R+3\Box\ln(\Omega)-\frac{3}{2}(\partial\ln(\Omega))^2 \right).
\end{aligned}
\end{equation}
The second term in $\ref{rtransform}$ will result in a boundary term in the Lagrangian. If we discard this term, the Lagrangian in the Einstein frame is given by:
   \begin{equation}
  \begin{aligned}
   \mathcal{L} &= \sqrt{-g}\left(\frac{R}{2}-\left(\frac{K}{\Omega}+\frac{3}{2}\frac{\Omega^{'2}}{\Omega^2}\right)\frac{1}{2}(\partial\phi)^2 -\frac{U}{\Omega^2}\right).
   \end{aligned}
 \end{equation}

 A field redefinition $\chi = \chi(\phi)$ can be done such that the kinetic energy term is canonically normalised:
    \begin{equation}\label{redefinition}
  \begin{aligned}
  \left(\frac{\delta \chi}{\delta \phi}\right)^2 &= \frac{K}{\Omega}+\frac{3}{2}\frac{\Omega^{'2}}{\Omega^2}.
  \end{aligned}
 \end{equation}
 
 In terms of the field $\chi(\phi)$, for which the explicit expression can be found by soling (\ref{redefinition}), the Lagrangian takes a simple form
 \begin{equation}\label{eframe}
  \begin{aligned}
   \mathcal{L} &= \sqrt{-g}\left(\frac{R}{2}-\frac{1}{2}(\partial \chi)^2-V(\chi) \right),
  \end{aligned}
 \end{equation}
where $V(\chi)$ is given by 
\begin{equation}\label{epotential}
V(\chi) = \frac{U(\phi(\chi))}{\Omega^2(\phi(\chi))}.
\end{equation}

\subsection{Slow-roll parameters in the Einstein frame}\label{slowroll}
In this subsection we express the Einstein frame potential slow-roll parameters in terms of the Jordan frame quantities $U$ and $\Omega$.  We only use potential slow-roll parameters\footnote{Since we only use potential slow-roll parameters, we will write $\epsilon$ and $\eta$ instead of $\epsilon_V$ and $\eta_V$ for the first and second potential slow-roll parameters}. The potential slow-roll parameters $\epsilon$ and $\eta$ in terms of the Einstein frame potential (\ref{epotential}) are given by\footnote{In the expressions of slow-roll parameters, the reduced Planck mass has been set to one (e.g. $\epsilon = \frac{m^2_{pl}}{16\pi}\left(\frac{_{V,\chi}}{V}\right)^2\rightarrow \frac{1}{2}\left(\frac{_{V,\chi}}{V}\right)^2$) }:
\begin{equation}\label{psrp}
\begin{aligned}
 \epsilon &= \frac{1}{2}\left(\frac{V_{,\chi}}{V} \right)^2, \space \eta &&= \frac{V_{,\chi\chi}}{V}
\end{aligned}
\end{equation}
We can also express the slow-roll parameters in terms of the Jordan frame potential $U(\phi)$, the conformal factor $\Omega(\phi)$, using the relation between the Jordan frame potential and the Einstein frame potential (\ref{epotential}), and the field redefinition (\ref{redefinition}). The first order potential slow-roll parameter $\epsilon$ is given by:
 \begin{equation}\label{jepsilon}
  \begin{aligned}
   \epsilon &= \frac{1}{2}\left(\frac{V_{,\chi}}{V}\right)^2 \\
   &= \frac{1}{2}  \left(\frac{\delta \phi}{\delta \chi}\right)^2 \left(\frac{\frac{d}{d\phi}\left(\frac{U}{\Omega^2}\right)}{\left(\frac{U}{\Omega^2}\right)}\right)^2 \\
   &= 2\frac{1}{\frac{K}{\Omega}+\frac{3}{2}\frac{\Omega^{'2}}{\Omega^2}}\left(\frac{U'}{U}-2\frac{\Omega'}{\Omega}\right)^2.
  \end{aligned}
 \end{equation}
The second order potential slow-roll parameter $\eta$ is given by:
  \begin{equation}\label{jeta}
  \begin{aligned}
   \eta &= \frac{V_{,\chi\chi}}{V} \\
   &= \frac{1}{V}\frac{\delta \phi}{\delta \chi}\frac{\partial}{\partial \phi}\frac{\delta \phi}{\delta \chi}\frac{\partial}{\partial \phi}V \\
   &= \frac{\Omega^2}{U}\frac{1}{\sqrt{\frac{K}{\Omega}+\frac{3}{2}\frac{\Omega'^2}{\Omega^2}}}\frac{\partial}{\partial \phi}\frac{1}{\sqrt{\frac{K}{\Omega}+\frac{3}{2}\frac{\Omega'^2}{\Omega^2}}}\frac{\partial}{\partial \phi}\frac{U}{\Omega^2} \\
   &= \frac{1}{(\frac{K}{\Omega}+\frac{3}{2}\frac{\Omega'^2}{\Omega^2})}\left[\left(\frac{U''}{U}-4\frac{U'\Omega'}{U\Omega}-2\frac{\Omega''}{\Omega}+6\frac{\Omega'^2}{\Omega^2}\right)\right] \\
   &+ \frac{1}{(\frac{K}{\Omega}+\frac{3}{2}\frac{\Omega'^2}{\Omega^2})^2}\left[\left(\frac{U'}{U}-2\frac{\Omega'}{\Omega} \right)\left(-\frac{1}{2}\right)\left( \frac{K'}{\Omega}-\frac{K}{\Omega}\frac{\Omega'}{\Omega}+3\frac{\Omega''}{\Omega}\frac{\Omega'}{\Omega}-3\frac{\Omega'^3}{\Omega^3}\right) \right], 
  \end{aligned}
 \end{equation}
where primes denote derivatives with respect to $\phi$. \\
Related to our expression for $\epsilon$, we find for the number of e-folds $N$:
\begin{equation}\label{jN}
 \begin{aligned}
  N &= \int_{\chi_{\text{end}}}^{\chi_N}d\chi\frac{V}{V_{,\chi}} \\
  &= \int_{\phi_{\text{end}}}^{\phi_N}d\phi \left(\frac{\delta \chi}{\delta \phi}\right)^2\frac{V}{V_{,\phi}} \\
  &= \int_{\phi_{\text{end}}}^{\phi_N}d\phi \left(\frac{K(\phi)}{\Omega(\phi)}+\frac{3}{2}\frac{\Omega'^2(\phi)}{\Omega^2(\phi)}\right)\frac{1}{\left(\frac{U'}{U}-2\frac{\Omega'}{\Omega}\right)} \\
 \end{aligned}
\end{equation}  
   
\subsection{Expansion of the Einstein frame potential}\label{vfullexpansion}
For the asymptotic Starobinsky model we have the Jordan frame potential $U = \lambda (\Omega-1)^2$. The Einstein frame potential $V$ is given by (\ref{epotential})
\begin{equation}
V(\chi) = \frac{U(\phi(\chi))}{\Omega^2(\phi(\chi))}.
\end{equation}
where $\chi$ is the canonically normalised field defined by the field redefinition (\ref{redefinitionk1})
    \begin{equation}
  \begin{aligned}
  \left(\frac{\delta \chi}{\delta \phi}\right)^2 &= \frac{1}{\Omega}+\frac{3}{2}\frac{\Omega^{'2}}{\Omega^2}.
  \end{aligned}
 \end{equation}
These two relations allow us to expand $V$ in terms of $\chi$ around the vacuum value $\chi_{\text{vac}}=0$:
\begin{equation}
 \begin{aligned}
  V(\chi) &= \sum_n c_n \chi^n,
 \end{aligned}
\end{equation}
where $c_n = \frac{1}{n!}\left.\frac{\delta^n V}{\delta\chi^n}\right|_{\chi_{\text{vac}}}$. Using the field redefinition (\ref{redefinitionk1}) we find:
\begin{equation}
\begin{aligned}
 c_0 &= 0 \\
 c_1 &= 0 \\
 c_2 &= \frac{\lambda \Omega'^2_{\text{vac}}}{1+\frac{3}{2}\Omega'^2_{\text{vac}}} \\
 c_3 &= \lambda \frac{\left(\Omega''_{\text{vac}}\Omega'_{\text{vac}}-\frac{3}{2}\Omega'^3_{\text{vac}} -\frac{3}{2}\Omega'^5_{\text{vac}}\right)}{\left(1+\frac{3}{2}\Omega'^2_{\text{vac}}\right)^{\frac{5}{2}}} \\
 c_4 &= \frac{1}{\left(1+\frac{3}{2}\Omega'^2_{\text{vac}}\right)^4}\frac{\lambda}{48} \begin{bmatrix}
 12 \Omega_{\text{vac}}''^2+63 \Omega_{\text{vac}}'^8+126 \Omega_{\text{vac}}'^6+24 \Omega_{\text{vac}}^{(3)} \Omega_{\text{vac}}'^3+16 \Omega_{\text{vac}}^{(3)} \Omega_{\text{vac}}'+ \\ \Omega_{\text{vac}}'^4 (75-36 \Omega_{\text{vac}}'')-4 \Omega_{\text{vac}}'^2 \Omega_{\text{vac}}'' (24 \Omega_{\text{vac}}''+25)\end{bmatrix}
\end{aligned}
\end{equation}
For $\Omega'^2_{\text{vac}} \ll 1$ we therefore obtain
\begin{equation}
 \begin{aligned}
  V(\chi) &\approx \lambda \Omega'^2_{\text{vac}} \chi^2+\lambda \left(\Omega''_{\text{vac}}\Omega'_{\text{vac}}-\frac{3}{2}\Omega'^3_{\text{vac}} \right)\chi^3 + \\
  &+ \frac{\lambda}{24} \left(-50\Omega'^2_{\text{vac}}\Omega''_{\text{vac}}+\frac{75}{2}\Omega'^4+6\Omega''^2_{\text{vac}}+8\Omega'_{\text{vac}}\Omega'''_{\text{vac}}\right)\chi^4+\dots
 \end{aligned}
\end{equation}
   
\subsection{Generalised asymptotic Starobinsky models}\label{sec:gs}
We consider the Jordan frame potential $U = \lambda\Omega^2(1-\Omega^{-1})^{2n}$. In the steep conformal limit this potential corresponds to the Einstein frame potential $V$ in terms of the canonically normalised field $\chi$
\begin{equation}
 \begin{aligned}
  V(\chi) &= \lambda \left(1-e^{-\sqrt{\frac{2}{3}}\chi} \right)^{2n},
 \end{aligned}
\end{equation}
which is called a generalised Starobinsky model. \\
 
The Einstein frame slow-roll parameters and the number of e-folds $N$ can be expressed in terms of $\Omega$ and its derivatives: 
 \begin{subequations}
 \begin{align}
  \epsilon &= \frac{2n^2}{\Omega+\frac{3}{2}\Omega'^2}\left(\frac{\Omega'}{\Omega-1}\right)^2 \label{gsepsilon}\\  
  \eta &= \frac{2n(2n-1)}{\Omega+\frac{3}{2}\Omega'^2}\left(\frac{\Omega'}{\Omega-1}\right)^2 
  + \frac{2n}{\left(\Omega+\frac{3}{2}\Omega'^2\right)^2}\frac{\Omega'}{\Omega-1}\left(-\frac{3}{2}\Omega'\Omega -\frac{3}{2}\Omega'^3+\frac{\Omega''}{\Omega'}\Omega^2\right) \label{gseta}\\
  N&= \frac{1}{n}\int^{\phi_N} d\phi \left(1+\frac{3}{2}\frac{\Omega'^2}{\Omega}\right)\left(\frac{\Omega-1}{2\Omega'}\right) \label{gsN}
 \end{align}
\end{subequations} 
\subsubsection*{Flat limit}
In the flat limit $1 \gg \frac{3}{2}\frac{\Omega'^2}{\Omega}$, (\ref{gsepsilon}), (\ref{gseta}) and (\ref{gsN}) simplify to: 
\begin{subequations}
\begin{flalign}   
  \epsilon &\approx 2n^2 \frac{\Omega'^2}{\Omega}\frac{1}{(\Omega-1)^2} \label{eweakk1} \\
  \eta &\approx 2n(2n-1)\frac{1}{(\Omega-1)^2}\frac{\Omega'^2}{\Omega}+2n\frac{1}{\Omega-1}\left(\Omega'' -\frac{3}{2}\frac{\Omega'^2}{\Omega} \right) \label{etaweakk1} \\
N &\approx \frac{1}{n}\int^{\phi_N}_{\phi_{\text{end}}}\frac{\Omega -1}{2\Omega'}d\phi \label{Nweakk1}
\end{flalign}
\end{subequations}
\subsubsection*{Flat conformal factor condition}
If $\Omega$ satisfies the flat conformal factor condition (\ref{finetuning}), $\Omega$ has the following form, for $\tilde{\phi} = \phi-\phi_{\text{vac}}$:
\begin{equation}
\begin{aligned}
 \Omega &= 1+\sum_{m=1} \Omega_m \alpha^m \tilde{\phi}^m,
\end{aligned}
\end{equation}
where the $\Omega_m$ are of order $O(1)$ and $\alpha \tilde{\phi} \ll 1$ for $0 < \tilde{\phi} < \tilde{\phi}_N$. This form allows us to evaluate (\ref{gsepsilon},\ref{gseta},\ref{gsN}) in orders of $\alpha$.  \\
\begin{exmp}\label{o1o2exp}
\textbf{$\Omega_1 \neq 0, \Omega_2 \neq 0$:}
\begin{subequations}\label{gso1o2}
\begin{flalign}
  \epsilon &= \frac{2n^2}{\tilde{\phi}^2}+(\alpha \tilde{\phi})\frac{2n^2}{\tilde{\phi}^2}\left(2\frac{\Omega_2}{\Omega_1}-\Omega_1\right)+O\left((\alpha\tilde{\phi})^2\right)  \label{gsepp} \\
  \eta &= \frac{2n(2n-1)}{\tilde{\phi}^2} +(\alpha \tilde{\phi})\frac{2n}{\tilde{\phi}^2}\left(4n\frac{\Omega_2}{\Omega_1}-(2n+\frac{1}{2})\Omega_1\right)+O((\alpha\tilde{\phi})^2) \label{gsetp} \\
  N &= \frac{\tilde{\phi}_N^2}{4n}-\frac{(\alpha \tilde{\phi}_N)}{6n}\frac{\Omega_2}{\Omega_1}\tilde{\phi}^2_N+O\left((\alpha\tilde{\phi}_N)^2\right)
\end{flalign}
\end{subequations}
This example illustrates that for generic $\Omega$ with non vanishing $\Omega_1$, the flat conformal factor attractor point of generalised asymptotic Starobinsky model of degree $2n$ corresponds to $\phi^{2n}$-chaotic inflation.
\end{exmp}

\begin{exmp}
 \textbf{$\Omega_1=\Omega_3 =0$, $\Omega_2\neq 0, \Omega_4\neq 0$} \\ 
If we impose $\Omega$ to be an even function of $\tilde{\phi}$, we find $\Omega_i=0$ for $i$ odd.
 \begin{subequations}\label{symmetric}
\begin{flalign}
  \epsilon &= \frac{2(2n)^2}{\tilde{\phi}^2}+(\alpha \tilde{\phi})^2\frac{2(2n)^2}{\tilde{\phi}^2}\left(2\frac{\Omega_4}{\Omega_2}-\Omega_2\right)+O\left((\alpha\tilde{\phi})^4\right)   \\
  \eta &= \frac{4n(4n-1)}{\tilde{\phi}^2} +\alpha^2\left(\frac{\Omega_4}{\Omega_2}4n(8n+1)-\Omega_24n(4n+1)\right) +O\left((\alpha\tilde{\phi})^4\right)\\
  N &= \frac{\tilde{\phi}^2_N}{8n}-\frac{1}{16n}\frac{\Omega_4}{\Omega_2}(\alpha\tilde{\phi}_N)^2\tilde{\phi}^2_N+O\left((\alpha\tilde{\phi})^4\right)
\end{flalign}
\end{subequations} 
\end{exmp}

\begin{exmp}
\textbf{$\Omega_1 = m, \Omega_2 = \binom{m}{2} $ (induced inflation with monomial)}
\\ In example (\ref{exmp1}) we have shown that we can expand the monomial induced inflation model $\Omega = \xi \phi^m$ in terms of $\tilde{\phi}= \phi-\phi_{\text{vac}}$ with $\Omega_1 = m, \Omega_2 = \binom{m}{2} $. For the generalised asymptotic Starobinsky (\ref{npotential}) model we find:
\begin{subequations}
\begin{flalign}
  \epsilon &= \frac{2n^2}{\tilde{\phi}^2}-(\alpha \tilde{\phi})\frac{2n^2}{\tilde{\phi}^2}+O((\alpha\tilde{\phi})^2)   \\
  \eta &= \frac{2n(2n-1)}{\tilde{\phi}^2} -(\alpha \tilde{\phi})\frac{2n^2}{\tilde{\phi}^2}\left(2+\frac{1}{2}\frac{m}{n}\right)+O((\alpha\tilde{\phi})^2) \\
  N &= \frac{\tilde{\phi}^2_N}{4n}-\frac{(m-1)}{12 n}(\alpha \tilde{\phi}_N ) \tilde{\phi}^2_N+O\left((\alpha \tilde{\phi}_N)^2 \right) 
\end{flalign}
\end{subequations} 
\end{exmp}

\begin{exmp}\label{uaplus}
\textbf{$\Omega = 1 + \alpha \phi^m$ (universal attractor inflation with monomial)}
\\ Universal attractor inflation \cite{Kallosh:2013tua} is a class of asymptotic Starobinsky models with $\Omega(\phi) = 1+\xi f(\phi)$. We consider the monomial case $f(\phi) = \phi^m$, for which $\phi_{\text{vac}}=0$ and extend the discussion to \emph{generalised} asymptotic Starobinsky models (\ref{npotential}). We find:
\begin{subequations}
\begin{flalign}
  \epsilon &= \frac{2(nm)^2}{\phi^2}-(\alpha\phi^m)\frac{2(nm)^2}{\phi^2}+O(\alpha^2\phi^{2m}) \\
  \eta &= \frac{2(nm)(nm-1)}{\phi^2}-(\alpha\phi^m)\frac{m^2n(4n+1)}{\phi^2}+O((\alpha^2\phi^{2m})) \\
  N &= \frac{\phi_N^2}{4mn}+ O((\alpha^2\phi^{2m})).
\end{flalign}
\end{subequations} 
These results correspond to leading order to $\phi^{2mn}$-chaotic inflation. This illustrates that if $\Omega$ satisfies the flat conformal factor condition (\ref{finetuning}), but with the first $m-1$ derivatives vanishing, the generalised asymptotic Starobinsky model of order $2n$ will lead to $\phi^{2mn}$-chaotic inflation, to leading order. 
\end{exmp}

\subsection{Beyond the asymptotic Starobinsky paradigm}
If $\Omega$ satisfies the flat conformal factor condition (\ref{finetuning}), the generalised asymptotic Starobinsky models (\ref{npotential}) have a chaotic inflation attractor point. We could also consider the much more general class of models satisfying that $U\rightarrow 0$ and $\Omega\rightarrow 1$ in the vacuum configuration:
\begin{equation}
 \begin{aligned}
  U &= \sum_{n\neq 0}a_n \left(\Omega -1\right)^n,
 \end{aligned}
\end{equation}
for a set of coefficients $\{a_n\}$. One can check that these general potentials are \emph{not} Starobinsky-like in the steep limit (\ref{strong}). They \emph{can} correspond to chaotic inflation if $\Omega$ satisfies the flat conformal factor condition (\ref{finetuning}). If $a_n = 0$ for $n<m$ and $a_m\neq 0$, than these models correspond to leading order to $\phi^{km}$-chaotic inflation, where $k$ is the first nonzero derivative of $\Omega$ in the vacuum.

\end{document}